\documentclass[
reprint,a4paper,twocolumn,aps,prl,amsmath,amssymb,floatfix,
superscriptaddress,showkeys,longbibliography
]{revtex4-2}

\usepackage[utf8]{inputenc}
\usepackage[T1]{fontenc}
\usepackage{microtype}

\usepackage{mathtools}   
\DeclareMathOperator{\Tr}{Tr}

\usepackage{graphicx}
\usepackage[usenames,dvipsnames,svgnames,table]{xcolor}

\usepackage{siunitx}
\usepackage{enumitem}
\usepackage{comment}
\usepackage{placeins}
\usepackage{soul}
\usepackage{xspace}
\usepackage{xifthen}

\usepackage[colorlinks,linkcolor=blue,citecolor=blue,urlcolor=blue]{hyperref}

\def\equationautorefname~#1\null{Eq.~(#1)\null}

\newcommand{\op}[1]{\hat{#1}}

\newcommand{\affilUKON}{Department of Physics, University of Konstanz, 78457 Konstanz, Germany.}

\begin{document}
	
	\title{Manifestations of flow topology in a quantum driven-dissipative system}
	
	\author{Kilian Seibold}\affiliation{\affilUKON}
	\author{Greta Villa}\affiliation{\affilUKON}
	\author{Javier del Pino}\affiliation{\affilUKON}\
	\affiliation{Departamento de Física Teórica de la Materia Condensada and Condensed Matter Physics Center (IFIMAC),
		Universidad Autónoma de Madrid, E-28049 Madrid, Spain}
	\author{Oded Zilberberg}\affiliation{\affilUKON}
	
	\date{\today}
	
	\begin{abstract}
		In driven-dissipative bosonic systems, the interplay between coherent driving, inter-particle interactions and dissipation leads to a rich variety of non-equilibrium stationary states (NESS).
		In the semiclassical limit, the \textit{flow topology} of phase-space dynamics governs the stability and structure of these dynamical phases.
		Consequently, topological transitions occur when the number of NESS, their chirality, or their connectivity changes, reflecting global reorganization in the system's dynamical phase-space landscape. 
		Here, we study the corresponding topological signatures in a driven-dissipative quantum Kerr oscillator. Employing a Lindblad master equation and quantum trajectory methods, we reveal that quantum dynamics retain key topological features of the underlying classical flows, with clear signatures accessible via quantum state tomography and linear response.
		In this manner, we predict new phases that are not signaled by Liouvillian gap closing, thereby generalizing the conventional criteria for diagnosing phase transitions.
		Our findings position phase-space flow topology as a powerful tool to identify and control robust quantum phases, enabling advances in error correction and sensing.
	\end{abstract}
	\maketitle
	
	Phase transitions involve abrupt changes in macroscopic behavior, often marked by order parameters that break a symmetry of the system.  
	In equilibrium at zero temperature, Ginzburg-Landau mean-field theories describe such transitions via energy functionals whose minima define the symmetry-broken phases~\cite{Hertz1976,Sachdev2011}. 
	This framework underpins our understanding of, e.g.,  ferromagnets~\cite{Schinz1992,Dahl2007}, superconductors~\cite{SaDeMelo1993,Schmidt2013,Fitzpatrick2017}, and Bose-Einstein condensates~\cite{Sieberer2013}, where local observables like magnetization act as order parameters. 
	Some phases, however, evade description by any local order parameters and are instead distinguished by global topological invariants. Topological insulators, topological superconductors, and various non-Hermitian systems are paradigmatic examples classified by topological indices~\cite{Schnyder2009,Hasan2010,Bernevig2013,Bergholtz2021}.
	Topological phase transitions occur when these invariants change abruptly, often signaled by a spectral gap closing or an abrupt change in the response of the system~\cite{Hasan2010, frankel2011geometry, Dalibard2011, Bernevig2013, Ozawa2019,porras_topological_2019,wanjura_topological_2020,Bergholtz2021,wan_quantum-squeezing-induced_2023}. 
	Independent of the underlying mechanism driving the phase transition, thermal and quantum fluctuations may enable the system to recover symmetry and ergodicity in the long-time limit~\cite{Sachdev2011}. 
	
	Dissipative phase transitions (DPTs) extend the concept of equilibrium phase transitions to open quantum systems~\cite{Bartolo2016,Fitzpatrick2017,Minganti2018,Soriente2021,Minganti2023,Beaulieu2025}. 
	Unlike isolated systems, which settle into equilibrium at potential minima, open systems instead relax into non-equilibrium steady states (NESS), where conservative and non-conservative forces balance asymptotically in the long-time limit~\cite{Esposito2009, Rivas2012}.
	Here, too, thermal and quantum fluctuations enable the system to explore the manifold of NESS, often rendering phase transitions as smooth crossovers accompanied by long-lived metastable states and critical slowing down~\cite{heugel2023role, heugel2024proliferation, Margiani2025}. As true non-analytic behavior emerges only in the limit of large excitations~\cite{Minganti2018, Vicentini2018}, 
	a commonly used tool to characterize DPTs is the closing of the Liouvillian gap, analogous to gap closings in Hamiltonian systems~\cite{Diehl2008, Minganti2018}.
	DPTs manifest in many systems, including
	Kerr parametric oscillators~\cite{Soriente2021,eichler2023classical, ChavezCarlos2024, Iyama2024},
	Bose-Hubbard lattices~\cite{Tomita2017,Vicentini2018},
	exciton-polariton condensates~\cite{Carusotto2013,Fink2018}, 
	superconducting qubit arrays~\cite{Flurin2017,Wang2019,Burkhart2021,Cai2021,Peyruchat2024,Kollar2024},
	driven-dissipative spin models~\cite{Kessler2012,Wang2021,Minganti2021}, and nonlinear photonic lattices~\cite{Verstraelen2020,Minganti2023}.

	Recently, the phase-space vector flows associated with the semiclassical equations of motion of driven resonators have been classified 
	by a topological graph index~\cite{villa2024}. The index captures how regions of phase space are dynamically connected~\cite{Oshemkov1998}, while also providing a comprehensive account of the topology of non-Hermitian excitations around each NESS. Crucially, between regions associated with different graph indexes, a flow topology transition (FTT) must occur, leading to a (global) change in the variety of NESS and their distribution in phase space~\cite{villa2024}. Such flow transitions have been confirmed through full graph reconstruction in a classical resonator, raising the question of whether the underlying topology also appears in more accessible system responses. 
	In the quantum regime, expectation values of observables generally follow similar flows, but intrinsic zero-point fluctuations introduce stochasticity that removes the notion of a deterministic vector flow in the system~\cite{Dalibard1992, Carmichael1993}.
	This raises a key question: do flow-topological features leave observable imprints in the quantum regime?

	Here, we reveal the  signatures of the flow topology in a driven-dissipative quantum Kerr resonator. This enables us to go beyond conventional Liouvillian gap analysis and uncover new phases.
	We propose clear linear-response observables sensitive to the flow topology  manifestations, and demonstrate their persistence under strong quantum fluctuations.
	Our findings establish a  framework for topologically characterizing quantum states in driven-dissipative systems, with implications for quantum error correction~\cite{Reinhold2020,Kwon2022}, quantum information~\cite{Slussarenko2019}, quantum sensing~\cite{Cai2021}, and quantum control~\cite{Dong2010}.

\begin{figure}[t!]
		\centering
		\includegraphics[width = 1.0\linewidth]{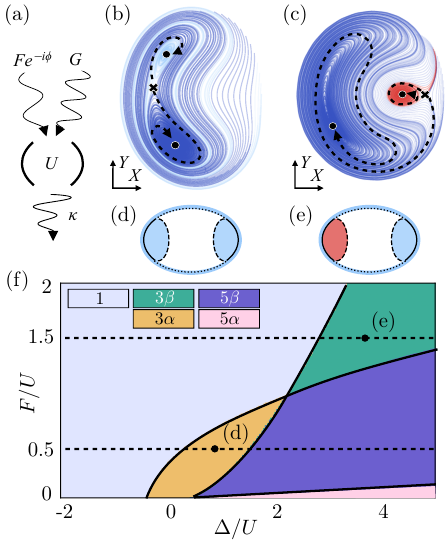}
		\caption{\textit{Semiclassical phase-space topology in a driven-dissipative resonator}~\cite{villa2024}.
			(a) Schematic of the driven-dissipative bosonic mode [cf. Eqs.~\eqref{eq:Hamiltonian RWA}-\eqref{eq:GPE}]. 
			(b) Phase-space vector flow defined by Eq.~\eqref{eq:GPE} evaluated at $(\Delta,F)/U = (0.7,0.5)$.  Crosses mark saddles; dots mark attractors, each corresponding to a deterministic NESS. Dashed arrows delineate separatrices, simplified for visualisation purposes.
			(c) Same as (b) at, $(\Delta,F)/U = (3.3,1.5)$. 
			(d) Topological graph index corresponding to (c), capturing the structure of two attractors (regions encircled by solid and dashed lines) separated by a saddle (region encircled by dotted and dashed lines). The (CW) chirality of the attractors associated with the (particle-like) character of local Bogoliubov excitations is marked by a blue face color.
			(e) Same as (d), corresponding to (c), with the CCW attractor (hole-like excitations) marked by red face color.
			(f) Topological phase diagram, where colors mark different topological variety of NESSs (flow pattern) corresponding to different graph indices.  
			Common parameters are: $\omega_0 = 10^2$, $U = 1$, $G = 0.4$, $\kappa = 0.1$, $\phi = 0$. 
		}
		\label{fig:1}
	\end{figure}

	
	We study a driven-dissipative Kerr resonator described by the effective Hamiltonian [Fig.~\ref{fig:1}(a)]
	\begin{align}
		\hat{\mathcal{H}} =& (-\Delta +U)\, \hat{b}^{\dagger}\hat{b} 
		+ \frac{U}{2} \hat{b}^{\dagger 2}\hat{b}^2+ \frac{G}{2}\left(\hat{b}^{\dagger 2} + \hat{b}^2\right) + \nonumber \\&F\left(\hat{b}^{\dagger} e^{-i\phi} + \hat{b} e^{i\phi}\right)\,,
		\label{eq:Hamiltonian RWA}
	\end{align}
	where $\hbar=1$ and $\hat{b}$ is the bosonic annihilation operator for drive photons~\cite{Kosata2022,seibold2025}. 
	The model \eqref{eq:Hamiltonian RWA} is written in a frame rotating at frequency $\omega$; the detuning between the drive frequency $\omega$ and the resonator frequency $\omega_0$ is  $\Delta=(\omega^2-\omega_0^2)/2\omega$.
	Here, $U$ is the Kerr nonlinearity, $G$ is the two-photon (parametric) drive amplitude at frequency $\omega_p=2\omega$, and $F$ is the amplitude of a single-photon coherent drive at frequency $\omega$, with phase $\phi$. 
	To account for dissipation, we model single-photon losses using a Lindblad master equation, where the system's density matrix $\hat{\rho}$ evolves under the Liouvillian $\mathcal{L}$~\cite{Breuer2007}
	\begin{equation}
		\dfrac{d\op{\rho}}{dt} =\mathcal{L}\hat{\rho} =-i[\hat{\mathcal{H}},\op{\rho}] + \kappa\left( \op{b}\op{\rho}\op{b}^\dagger - \frac{1}{2}(\op{b}^\dagger\op{b}\op{\rho}+\op{\rho}\op{b}^\dagger\op{b})\right).
		\label{eq:quantum master equation}
	\end{equation} 

	\begin{figure*}[t]
		\centering
		\includegraphics[width = 1.0\linewidth]{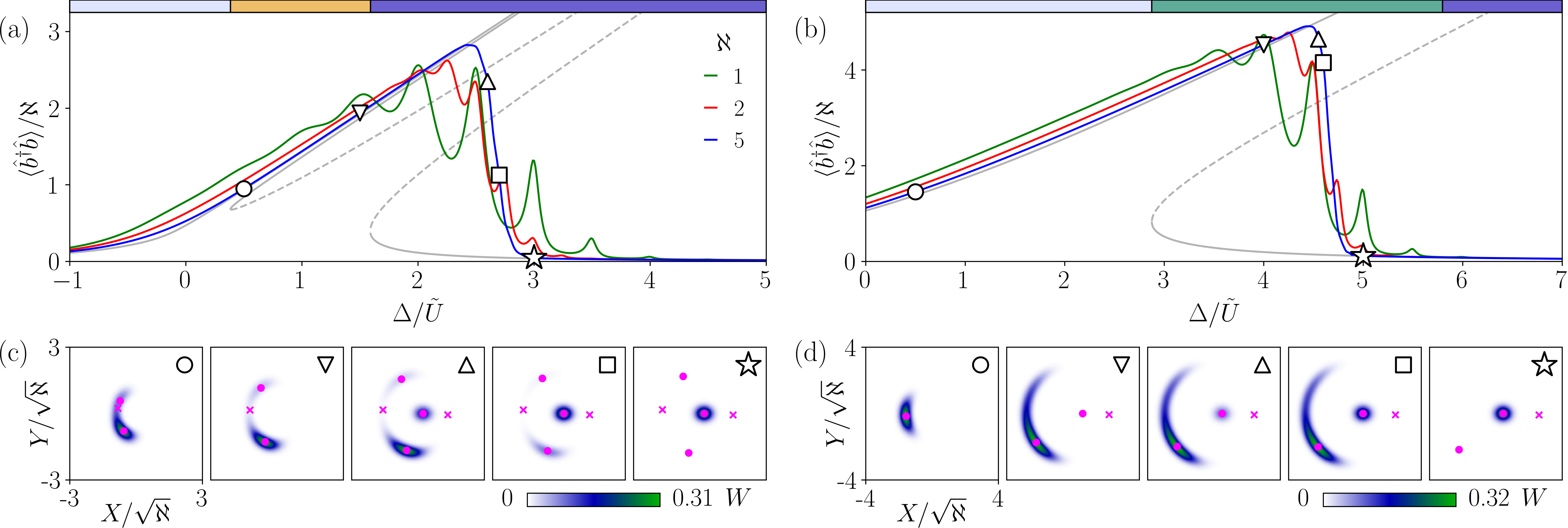}
		\caption{\textit{Classical versus quantum stationary states across topological flow transitions}. (a) and (b) Resonator populations (RSSP) along $F = 0.5$ and $F = 1.5$ cuts of Fig.~\ref{fig:1}, rescaled by $\aleph$ as quantum fluctuation levels vary. Gray solid (dashed) lines mark the classical population, $|\beta_0^k|^2$ associated with the stable (unstable) FPs of the corresponding Eq.~\eqref{eq:GPE}. Colored stripe indicates the corresponding classical phases as shown in Fig.~\ref{fig:1}(f). (c) and (d) Wigner functions for $\aleph = 5$ associated respectively to  (a) and (b) at representative detuning values, marked by symbols. Gray dots (crosses) mark stable (unstable) classical FPs.
		}
		\label{fig:2}
	\end{figure*}

	\textit{Flow-topology classification of classical dynamics} --- %
	In the semiclassical limit of large occupations, the driven-dissipative dynamics set by Eq.~\eqref{eq:quantum master equation} reduce to a deterministic evolution in phase-space.
	The dynamics then obey the Gross-Pitaevskii equation (GPE), a mean-field description of Eq.~\eqref{eq:quantum master equation} for the complex field $\beta=\mathrm{Tr}(\hat{\rho}\hat{b})$,
	\begin{equation}
		\begin{aligned}
			i\dfrac{d\beta}{d t} = &\left(- \Delta - i\frac{\kappa}{2} + U|\beta|^2\right)\beta + G\,\beta^* +\,F\,e^{-i\phi}\;.~\hspace{-4mm}
			\label{eq:GPE}
		\end{aligned}
	\end{equation} 
	This equation defines a two-dimensional vector flow in the rotating frame phase space $\{X, Y\} \equiv \{\mathrm{Re}(\beta), \mathrm{Im}(\beta)\}$, see, e.g., Figs.~\ref{fig:1}(b) and (c). 
	Stationary solutions correspond to the fixed points (FPs), $\beta_0^k$, found by setting $d\beta_0/dt = 0$ in Eq.~\eqref{eq:GPE} and solving for the roots of the resulting polynomial equations~\cite{ourPackage,Breiding_2018}.
	The flows organize around these FPs: stable ones are attractors (corresponding to NESS), with a local flow pattern that spirals clockwise (CW), counterclockwise (CCW), or not at all; unstable ones are saddles. Separatrices, crossing saddles, partition phase space into different basins of attraction.
	By studying the eigenspectrum of the dynamical matrix associated with Bogoliubov excitations around each FP, $\beta_0^k$, we can assess its stability, i.e., whether they are attractors or saddles, as well as their spiraling chirality~\cite{Soriente2021,Dumont2024a,villa2024}.
	Note that the vector flow defined by Eq.~\eqref{eq:GPE}  contains no repellors~\cite{Dumont2024a,villa2024}.
	%

	The flow structure governed by Eq.~\eqref{eq:GPE} admits a compact topological representation through a graph invariant, obtained by extending Morse–Smale theory~\cite{Oshemkov1998} to non-equilibrium flows~\cite{villa2024}, see Figs.~\ref{fig:1}(d) and (e)~\cite{supmat}.
	This invariant encodes (i) the number and stability of FPs, (ii) their connectivity, and (iii) local flow chirality around each NESS; The number of faces reflects (i), the network of edges between faces captures (ii), and face colors encode (iii)~\cite{villa2024}.
	Changes to the invariant under smooth parameter variations indicate topological obstructions to deformation, signaling phase transitions.
	This allows for the construction of a full topological phase diagram of structurally stable semiclassical flow patterns, see Fig.~\ref{fig:1}(f).
	
	The mean-field phase diagram in Fig.~\ref{fig:1}(f) illustrates this structure, mapping out regions where the flow invariant remains constant and thus defines a distinct dynamical phase~\cite{villa2024}.
	For example, region 1, with a single stable FP, cannot be smoothly deformed into regions $3\alpha$ and $3\beta$, each of which host three solutions—one saddle and two stable FPs—as shown in Figs.~\ref{fig:1}(b)-(e).
	Phases $3\alpha$ and $3\beta$ differ only in the chirality of the local flow around stable FPs: $3\alpha$ features clockwise (CW) rotation at both attractors, while $3\beta$ displays opposite chiralities~\cite{Dykman1980,Drummond1980}.
	They are separated by a saddle-node bifurcation at their phase boundary in Fig.~\ref{fig:1}(f).
	Similarly, the five-solution phases $5\alpha$ and $5\beta$ each contain three stable FPs (two CW, one CCW) and two saddles, but differ in how the CCW attractor connects to the saddles~\cite{villa2024, supmat}.
	
	\begin{figure*}[t]
		\centering
		\includegraphics[width = 1.0\linewidth]{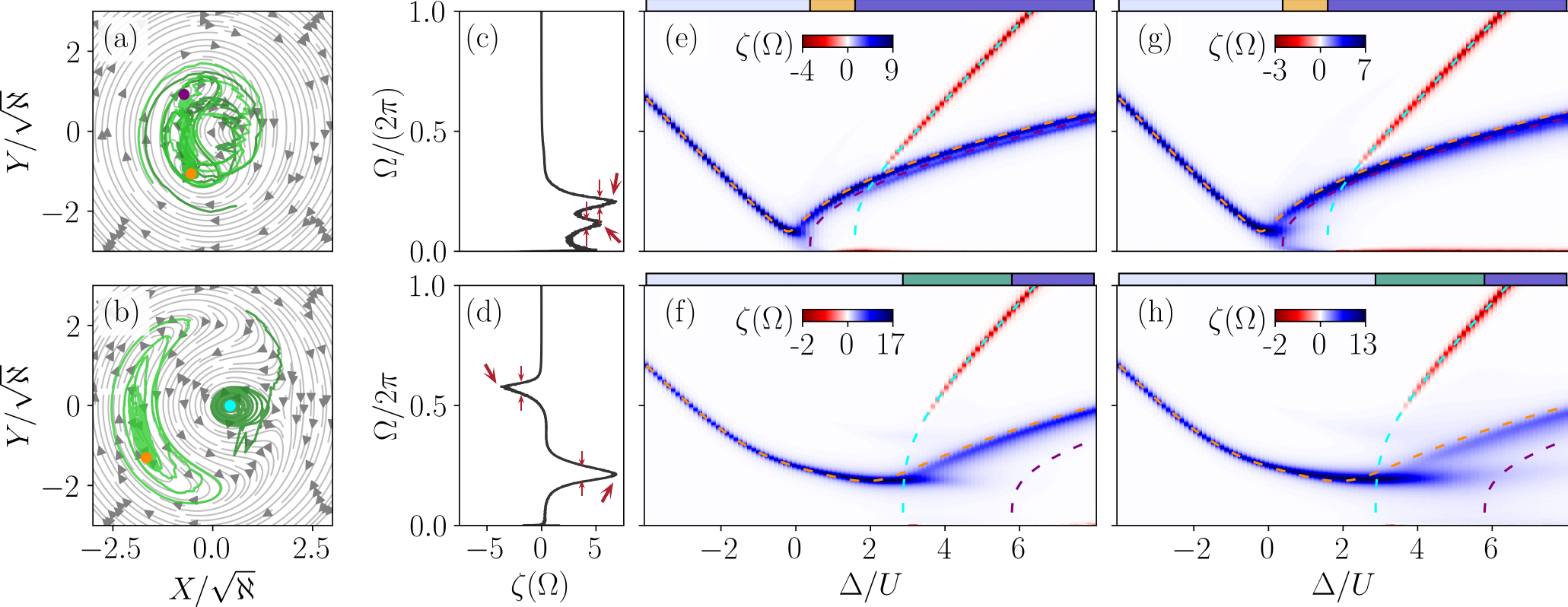}
		\caption{
			\textit{Chirality signatures extracted from quantum jump trajectories.}
			(a) Representative quantum jump trajectories (green) in region $3\alpha$  in the semiclassical regime ($\aleph=20$), for $(F,\Delta)/U=(0.5,1)$, overlaid on the corresponding classical flow (in gray) and attractors [dots colored as in the phase diagram in Fig.~\ref{fig:1}(e)]. (b) Same as (a) in region  $3\beta$ for $(F,\Delta)/U=(1.5,4)$.
			(c) and (d) Chirality spectrum $\zeta(\Omega)$ [cf.~Eq.~\eqref{eq:chirality spectrum}] for panels (a) and (b), respectively. Arrows mark the resonance peaks and their width. 
			(e) and (f) Chirality spectra for $\aleph=20$ along the detuning cuts in Fig.~\ref{fig:1}(f), cf.~Fig.~\ref{fig:2}.  We sweep across the dynamical phase transitions $1 \to 3\beta \to 5\beta$ and $1 \to 3\alpha \to 5\beta$, respectively. Colored stripes as in Fig.~\ref{fig:2}. 
			Dashed lines trace the expected semiclassical response frequency around each attractor with corresponding color to (a) and (b). 
			(g) and (h) Same cuts as (e) and (f), with stronger quantum fluctuations ($\aleph=10$). New features appear aside from those traced by dashed lines.
		}
		\label{fig:3}
	\end{figure*}

	\textit{Quantum NESS: ensemble averaging and topology }---
	We now examine whether and how these classical topological features persist in the fully quantum non-equilibrium steady state, beyond the mean-field approximation.
	To address this, we numerically compute the stationary density matrix $\hat{\rho}_0$ by diagonalizing the Liouvillian from Eq.~\eqref{eq:quantum master equation} in a truncated Hilbert space.
	To interpolate between the classical and quantum regimes, we rescale the nonlinearity and drive amplitudes according to 
	\begin{equation}
		\tilde{\beta} = \beta/\sqrt{\aleph}
		\quad,\quad
		U= \tilde{U}/\aleph \quad\text{and}\quad
		F=\tilde{F}\sqrt{\aleph},
	\end{equation}  
	where $\aleph$ is the scaling parameter. 
	This scaling transformation leaves the GPE~\eqref{eq:GPE} invariant~\cite{Seibold2020, Seibold2022}.
	Increasing $\aleph$ reduces quantum fluctuations, pushing the system toward the classical regime.
	In Figs.~\ref{fig:2}(a) and (b), we show the rescaled steady-state population (RSSP), $\mathrm{Tr}[\hat{b}^\dagger\hat{b}\hat{\rho}_0]/\aleph$, along cuts from Fig.~\ref{fig:1}(f), overlaid with the mean-field squared amplitudes $|\beta_0^k|^2$.
	The RSSP increases during a sweep up in detuning until a first-order dissipative phase transition occurs (second-order for $F=0$)~\cite{Bartolo2016,heugel2019quantum,eichler2023classical}, inducing an abrupt drop in population.
	Importantly, the location of this DPT in parameter space does not align with the phase boundary predicted by the classical phase-space dynamics. Instead, it lies deep within the bistable phases $3\alpha$ and $3\beta$, and for small $\aleph$, its position is further obscured by regions of multi-photon resonances.
	This discrepancy arises because the density matrix $\rho_0$ is distributed across phase space due to ergodicity. The RSSP in Figs.~\ref{fig:2}(a) and (b) therefore reflects a weighted average over all attractors.
	Such averaging reduces sensitivity to the features that define the semiclassical topological flows, and in the multi-photon regime, these features may be entirely washed out.
	Since these structures determine the classical topological organization, one may conclude that quantum NESS have limited resemblance to their classical counterparts.
	
	To restore sensitivity to the classical flow features that are obscured in the RSSP, we examine the Wigner function $W(X,Y)$, a quasi-probability phase-space representation of $\hat{\rho}_0$~\cite{Fox2013}.
	In the classical limit ($\aleph \gg 1$), $W$ reduces to an incoherent mixture of Gaussians centered at the stable FPs of Eq.~\eqref{eq:GPE}.
	At moderate quantum fluctuations ($\aleph = 5$), $W$ develops intricate patterns around the FPs, as shown in Figs.~\ref{fig:2}(c) and (d).
	For example, at $ \tilde{F}/\tilde{U} = 0.5$, a positive detuning sweep along $\Delta/\tilde{U} \geq 0$ through regions 1, $3\alpha$, and $5\beta$ in Fig.~\ref{fig:1}(a) transforms $\hat{\rho}_0$ from a low-amplitude coherent state into a squeezed state, then into a bimodal and into a trimodal distribution [cf.~Fig.~\ref{fig:2}(c)]. In the final stage, the stationary distribution concentrates around a single low-amplitude FP, even though the classical mean-field flow supports three stable FPs.
	A similar sequence occurs in Fig.~\ref{fig:2}(d) for a cut at $\tilde{F}/\tilde{U} = 1.5$, crossing regions 1, $3\beta$, and $5\beta$ in Fig.~\ref{fig:1}(a). Here, a bimodal distribution forms before quantum fluctuations eventually localize the state around the low-amplitude FP.
	Crucially, the phase-space structure of $W(X,Y)$ closely tracks the classical flow topology of Eq.~\eqref{eq:GPE}.
	The number of prominent lobes reflects the number of attractors, while their shape and tails trace classical flow lines and larger structures, including 
	basins of attraction and often extending into saddle regions.
	Thus, even in the presence of quantum mixing and fluctuations,  Wigner quasi-probabilities largely retain semiclassical topological features.

	\textit{Chirality sensitive response function}---%
	While phase-space distributions obtained from the full density matrix capture many topological features, they remain insensitive to the chirality around each FP, encoded in the face colors of the graph invariant.
	This insensitivity stems from the fact that the density matrix $\hat{\rho}_0$ pertains to an ensemble average over all quantum trajectories. 
	As a result, the Wigner function reflects this averaging and smoothens out the local chirality experienced by individual trajectories exploring distinct regions of phase space. To capture these features in an observable, we define the \emph{chirality spectrum} as the causal (retarded) response function
	\begin{equation}
		\label{eq:chirality spectrum}
		\zeta(\Omega) = \int_{0}^{\infty}\!e^{\,i\Omega \tau}\,
		\bigl\langle Y(\tau)X(0)-X(\tau)\,Y(0)\bigr\rangle_{\!\rm ss}\,d\tau\,,
	\end{equation}
	with steadystate expectation value $\langle \square \rangle_{\rm ss} = \Tr[\square\, \hat \rho_{0}]$.
	This construction, inspired by Hamiltonian reconstruction~\cite{Dumont2024a,Soriente2018}, captures how winding of trajectories around FPs distributes over fluctuation Fourier frequencies $\Omega$.
	Away from bifurcations, each non-degenerate fluctuation mode produces a peak in $\zeta(\Omega)$, so the number of peaks coincides with the number of stable FPs. Saddles do not contribute, since their chirality is ill-defined.
	Peak positions (linewidths) correspond to the imaginary (real) parts of the Bogoliubov eigenvalues near each FP, while their signs indicate CW (positive) or CCW (negative) quantum trajectory rotation~\cite{Dumont2024a,Soriente2018}.
	Thus, $\zeta(\Omega)$ compactly encodes the essential ingredients of the graph invariant: the number of stable FPs equals to the number of colored faces, and their chirality encodes the face colors.
	
	In order to evaluate the chirality spectrum~\eqref{eq:chirality spectrum}, we \textit{unravel} the Lindblad master equation~\eqref{eq:quantum master equation} into single-trajectory evolutions via the Monte Carlo wave-function (quantum jump) method~\cite{Dalibard1992, Carmichael1993, Dum1992a,Dum1992,Moelmer1993}.
	In this approach, each wave function $\lvert\psi_r(t)\rangle$ evolves according to a non-Hermitian Schrödinger-like equation:
	\begin{equation}
		\frac{d}{dt}\lvert\psi_r(t)\rangle = -i\,\big(\hat{\mathcal{H}} - \frac{i\kappa}{2}\,\hat b^\dagger \hat b\big)\,\lvert\psi_r(t)\rangle\;,
		\label{eq:schrodinger}
	\end{equation}
	until a random photon‐emission ``jump'' occurs with probability $
	dp = \kappa\,n(t)dt$, with $n(t)= \langle\psi_r(t)\lvert \hat b^\dagger\hat b\rvert\psi_r(t)\rangle$. The wave function is then projected onto $
	\lvert\psi_r(t+dt)\rangle = \hat b\,\lvert\psi_r(t)\rangle/(\sqrt{n(t)})$.
	A single realization corresponds to the  outcome of a continuously monitored experiment, reflecting real-time quantum jumps as detected in, e.g. circuit QED systems~\cite{Plenio1998, weber2014mapping,Naghiloo2017, Minev2019,jordan2024quantum}.
	Averaging over many ($\mathcal{N}$) such realizations, with trajectory label $r$, yields the density matrix: $\overline{\lvert\psi_r(t)\rangle\langle\psi_r(t)\rvert} \equiv (1/\mathcal{N})\sum_{r=1}^\mathcal{N} \lvert\psi_r(t)\rangle\langle\psi_r(t)\rvert = \hat{\rho}(t)$, matching the solution of Eq.~\eqref{eq:quantum master equation}.  
	%

	Unlike the stationary state $\hat{\rho}_0$, individual quantum trajectories, with phase space coordinates  $(X_r(t), Y_r(t))$ given by the real and imaginary parts of $\tilde{\beta}_r(t) = \langle\psi_r(t)|\hat{b}|\psi_r(t)\rangle$, retain the full topological information about the underlying flow. In the semiclassical regime (large $\aleph$), each quantum trajectory closely follows a classical streamline, drifting away from saddle points and fluctuating around stable FPs at long times, see Figs.~\ref{fig:3}(a) and (b).
	Therefore, the state $|\psi_r(t)\rangle$ witnesses flow topology in three ways: 
	(i) the number and stability of FPs is encoded in the relative residence times within each basin, 
	(ii) the local chirality appears in the circulation of stochastic loops around each FP, and 
	(iii) the connectivity of FPs emerges from rare stochastic jumps that carry a trajectory between basins.
	From this trajectory viewpoint, the chirality spectrum~\eqref{eq:chirality spectrum} can be written as
	\begin{equation}
		\label{eq:chirality spectrum trajectories}
		\zeta(\Omega) = -\frac{2}{\mathcal{N}}\sum_{r=1}^\mathcal{N} \mathrm{Im}[\mathcal{F}(G_r(\tau))],
	\end{equation}
	where on each trajectory we evaluate the time-delayed correlator $G_r = \langle Y_r(t) X_r(t+\tau) - X_r(t) Y_r(t+\tau) \rangle_t$, with $\tau>0$ and $\langle\square\rangle_t$ denotes an average over all times $t$. We, then, average over the imaginary part of the Fourier transform, $\mathcal{F}$, of the correlator.

	We compute $\zeta(\Omega)$ along the detuning sweeps of Fig.~\ref{fig:2}, in the semiclassical regime ($\aleph = 20$), covering regions 1, $3\alpha$, and $3\beta$, see Figs.~\ref{fig:3}(c)–(f). In region 1, we observe a single CW mode that softens near the $1 \to 3\alpha$ transition. In region $3\alpha$, two nearly degenerate CW peaks emerge, see Fig.~\ref{fig:3}(e). In region $3\beta$, one CW and one CCW peak of opposite sign appear, see Fig.~\ref{fig:3}(f).
	Finally, in region $5\beta$, the spectrum exhibits two CW peaks and one CCW peak, reflecting the three distinct attractors of the system.
	Deeper in the quantum regime ($\aleph = 10$), see Figs.~\ref{fig:3}(g) and (h), the distinction between $3\alpha$ and $3\beta$ persists away from phase boundaries, but their sharp transition smoothens into a crossover. At the same time, multiphoton resonances involving higher eigenstates contribute, cf.~Fig.~\ref{fig:2} and Refs.~\cite{Bartolo2016,roberts2020}.
	Interestingly, the CW and CCW winding of the $\zeta(\Omega)$ peaks survives in the multiphoton regions. However, the connection to the underlying classical flow weakens, and mixed spectral features from both $3\alpha$ and $3\beta$ can appear at fixed $\Delta/U$.

	In the deep quantum limit ($\aleph = 1$), quantum fluctuations strongly modify the classical phase-space structure: Wigner peaks no longer match semiclassical FPs, and trajectories tunnel between multiple classical basins. This makes the evaluation of the chirality spectrum~\eqref{eq:chirality spectrum} via quantum trajectories challenging.
	Interestingly, the chirality spectrum~\eqref{eq:chirality spectrum} can still be evaluated in this limit, even without well defined classical attractors, using the spectral decomposition of the master equation in Eq.~\eqref{eq:quantum master equation}; we recall that the state at long times is well-approximated by a small number of Liouvillian eigenmodes (complex eigenvalues $\lambda_k$) on top of the steady state $\hat{\rho}_0$ ($\lambda_0=0$)~\cite{Minganti2018}. 
	\begin{figure}[t]
		\centering
		\includegraphics[width = 1.0\linewidth]{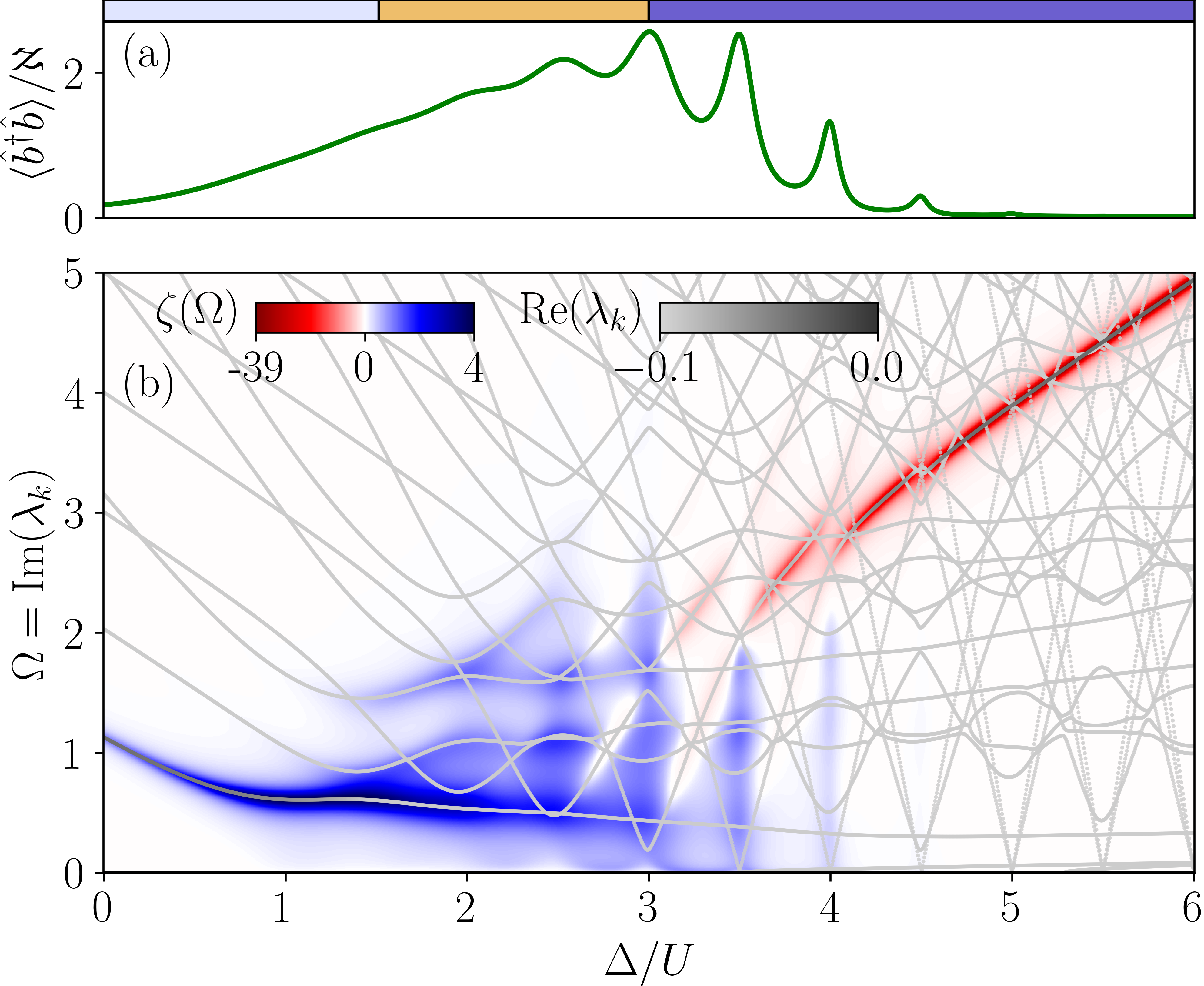}
		\caption{\textit{Chirality spectrum in the multiphoton resonance regime.} 
			(a) RSSP as a function of detuning $\Delta/U$ for $F=0.5$, $G=0.4$ in the deep quantum regime ($\aleph=1$). The top colored stripe marks the corresponding classical phases.
			(b) Background colormap: corresponding chirality spectrum $\zeta(\Omega)$ [cf.~Eq.~\eqref{eq:chirality spectrum}]. 
			Overlaid lines: imaginary parts $\mathrm{Im}(\lambda_k)$ of the Liouvillian eigenvalues, corresponding to coherent oscillatory modes; shading indicates decay rates $\mathrm{Re}(\lambda_k)$ (darker = slower decay).
			Even in the deep quantum regime, the spectral branches retain clear CW or CCW winding, allowing $\zeta(\Omega)$ to combine them across branches into a robust observable.
		}
		\label{fig: deep quantum chirality}
	\end{figure}
	By the quantum regression theorem~\cite{supmat}
	\begin{equation}
		\zeta(\Omega)=\sum_{k>0} \frac{w_k}{i\Omega-\lambda_k}\;,
		\label{eq:zeta_liouvillian}
	\end{equation}
	with $w_k=\Tr[\hat Y\,\hat r_k]\Tr[\hat \ell_k^\dagger \hat X\,\hat\rho_{0}]
	-\Tr[\hat X\,\hat r_k]\Tr[\hat \ell_k^\dagger \hat Y\,\hat \rho_{0}]$
	where $\hat{\ell}_k$ ($\hat{r}_k$) are the left (right) Liouvillian eigenvectors. 
	Equation~\eqref{eq:zeta_liouvillian} links each chirality peak to a Liouvillian mode with frequency $\mathrm{Im}(\lambda_k)$, linewidth $\mathrm{Re}(\lambda_k)$, and spectral weight set by overlap factors. 
	Unraveling $\hat{\ell}_k,\hat{r}_k$ into quantum trajectories, $\zeta(\Omega)$ encodes in each peak the \textit{net chirality} of the trajectory bundle forming it, which in the classical limit corresponds to trajectories confined within their basins of attraction, where traces thereof remain even in the deep quantum limit, see Fig.~\ref{fig: deep quantum chirality}.
	Moreover, sweeping $\Delta/\tilde{U}$ across multiphoton resonances [Fig.~\ref{fig: deep quantum chirality}(a)], the chirality spectrum $\zeta(\Omega)$ [colormap in Fig.~\ref{fig: deep quantum chirality}(b)] 
	reveals that chiralities can flip at avoided crossings ($\Delta/\tilde{U}=2$) or persist ($\Delta/\tilde{U}=3$), pointing to a reorganization of quantum trajectories in phase space. All these transitions occur without a Liouvillian gap closing, the standard hallmark of dissipative phase transitions~\cite{Minganti2018}. 
	Our results provide a mechanism to identify dissipative phase boundaries in the deep quantum regime through response functions, thereby transcending the conventional Liouvillian gap–closing criterion. A comprehensive analysis of the implications is left for future work

	\textit{Conclusion and Outlook}---%
	We have shown that classical flow topology leaves distinct and experimentally accessible fingerprints in driven-dissipative quantum steady states.
	In a prototypical Kerr resonator with single- and two-photon drives, the Wigner function traces central semiclassical flow-topological features, including the number, arrangement, and stability of classical FPs. Yet, ensemble averaging washes out the local chirality from the density matrix, encoded in individual trajectories.
	The frequency-resolved chirality spectrum~\eqref{eq:chirality spectrum} overcomes this limitation by extracting local winding signatures from fluctuation modes. Remarkably, these spectral features remain robust deep in the quantum regime, even when quantum fluctuations smear out classical attractors.
	All proposed diagnostics are directly accessible via heterodyne detection of the output field, providing a practical route to probe flow topology in current superconducting and photonic platforms~\cite{CampagneIbarcq2016, Blais2021,Cai2021}.
	Looking ahead, natural extensions include multimode generalizations, exploring the link between flow topology and photon blockade~\cite{roberts2020}, and exploiting topological signatures to develop bosonic codes tailored for fault-tolerant quantum computation~\cite{LabayMora2025, Guillaud2023}. More broadly, our results establish a spectral framework to characterize nonequilibrium transitions beyond the standard Liouvillian gap paradigm.

	\section*{Acknowledgements}
	We acknowledge funding from the Deutsche Forschungsgemeinschaft (DFG) through project numbers 449653034, 521530974, and 545605411, as well as via SFB 1432 (project number 425217212). We also acknowledge support from the Swiss National Science Foundation (SNSF) through the Sinergia Grant No. CRSII5\_206008/1. J.d.P. acknowledges support from the Ramón y Cajal program (RYC2023-043827-I).

	\bibliographystyle{apsrev4-2}
	\bibliography{refs}

	\FloatBarrier

\end{document}